# Dynamic elastic properties and magnetic susceptibility across the austenite-martensite transformation in site-disordered ferromagnetic Ni-Fe-Al alloy

P. K. Mukhopadhyay[1†a] and S. N. Kaul[2]

[1]LCMP, S. N. Bose National Center for Basic Sciences, Salt Lake, Kolkata – 700 098, India

[2]School of Physics, University of Hyderabad, Central University P. O., Hyderabad – 500 046, Andhra Pradesh, India

Besides permitting an accurate determination of the ferromagnetic-to-paramagnetic phase transition temperature and the characteristic temperatures for the beginning and end of the growth of martensite (austenite) phase at the expense of austenite (martensite) phase while cooling (heating), the results of an extensive ac susceptibility, sound velocity and internal friction investigation of the thermoelastic martensitic transformation in melt-quenched (site-disordered) $Ni_{55}Fe_{20}Al_{25}$ alloy provide a clear experimental evidence for the following. Irreversible thermoelastic changes (thermal hysteresis) occur in the austenite phase in the premartensitic regime. In the heating cycle, the system retains the "memory" of the initiation and subsequent growth of the martensitic phase (at the expense of the parent austenite phase) that had taken place during the cooling cycle in the austenite-martensite phase coexistence region. We report and discuss these novel findings in this communication.

[a] author to whom correspondences may be addressed (email: pkm@bose.res.in)

Thermoelastic martensitic transformation from face-centered-cubic (fcc) austenite (high-temperature) phase to tetragonal martensite (low-temperature) phase in a new ternary ferromagnetic alloy system Ni-Fe-Al ("prepared" in different states of site disorder) has been recently reported[1] based on the results of a detailed neutron diffraction (ND), electrical resistivity, $\rho(T)$, and magnetization, $M(T,H)$, studies. The effect of site disorder, in this alloy system, is to[1] (i) promote ductility, (ii) narrow down the temperature range over which austenite and martensite phases coexist (and hence sharpen the martensitic phase transition) and (iii) shift the martensitic transition temperature to higher temperatures. In the melt-quenched sample of $Ni_{55}Fe_{20}Al_{25}$ (henceforth referred to as $q$-$Fe_{20}$), which has the highest degree of site disorder, the thermoelastic martensitic phase transition is sharp and occurs near the Curie temperature $T_C \cong 225K$. The martensitic transformation (MT) is evidenced by the thermal hysteresis exhibited by electrical resistivity as a function of temperature, $\rho(T)$, when a given sample undergoes thermal cycling. An elaborate analysis of the $\rho(T)$ data yields[1] the characteristic temperatures for the beginning, $T_{Mb}$ ($T_{Ab}$), and end, $T_{Me}$ ($T_{Ae}$), of the growth of martensite (austenite) phase at the expense of austenite (martensite) phase while cooling (heating) as $T_{Mb} \cong 260K$, $T_{Me} \cong 150K$, $T_{Ab} \cong 170K$ and $T_{Ae} \cong 280K$ for the $q$-$Fe_{20}$ sample. These values for the characteristic temperatures are consistent with those deduced from the ND data.

Recognizing that the thermoelastic martensitic transformation causes the shape memory effect[2-7] and the thermoelastic nature of MT is directly reflected[4,5,8] in the elastic property measurements across the austenite-martensite phase transformation, extensive measurements of dynamic elastic properties, such as sound velocity and attenuation, of the sample $q$-$Fe_{20}$ were undertaken, using the vibrating reed technique[9-11]. This sample undergoes[1] a ferromagnetic (FM) to paramagnetic (PM) transition at $T_C \cong 225K$ (the Curie temperature), which falls within the temperature range over which MT occurs. Since the intrinsic magnetic susceptibility diverges at $T = T_C$ and is extremely sensitive to any structural changes, the real ($\chi_0'$) and imaginary ($\chi_0''$) components of ac susceptibility were measured as well.

Starting with 99.99 % pure *Ni*, *Fe* and *Al*, taken in proper proportions, polycrystalline rods of 10 mm diameter and 100 mm length with nominal composition $Ni_{55}Fe_{20}Al_{25}$ were prepared by radio-frequency induction-melting technique. A portion of these rods was melt-quenched to form 2 mm wide and $\approx 30\mu m$ thick crystalline ribbons. The details about the sample preparation and actual composition are furnished elsewhere[12]. Vibrating reed experiments were performed on an absolutely flat ribbon strip of length 25 mm, which forms the reed. In such experiments, one end of the reed is clamped firmly to the base plate of the cryostat while a flexural resonance is set up at the other (free) end by an electrostatic drive via an electrode placed near that end. Another matching electrode (biased at 45 volts and connected to the input of a lock-in amplifier), placed against the opposite face of the reed, picks up the oscillations by electrostatic coupling. The reference input to the lock-in amplifier comes from the same generator that feeds the drive electrode with sine wave amplitude variation, but locked to the second harmonic. At room temperature, the resonance curve is obtained by measuring the amplitude of vibration of the free end as a function of the driving frequency. From the resonance curve, so measured, the absolute value of internal friction, $Q^{-1} = \Delta\nu / \nu_{res}$, (where $\Delta\nu$ is the width of the resonance curve between the points at which the amplitude has the value $A_{\max} / \sqrt{2}$ if $A_{max}$ is the maximum amplitude at the resonance frequency $\nu_{res}$) at room temperature is deduced as $Q^{-1}(T = 293.7\,K) = 6.5\,x10^{-3}$. Subsequently, the resonance is phase-locked and tracked as the sample temperature is varied. The sound velocity, *V*, as a function of temperature is obtained from the fundamental (resonance) vibration frequency, $\nu$, using the relation $\nu = (d / 4\pi\sqrt{3})\,(1.875/l)^2\,V$, where *d* and *l* are the thickness and length of the sample reed. Note that we can measure only the relative changes in sound velocity, $\delta V / V$, with respect to its value at room temperature because the unknown clamping yield causes a large uncertainty in the measurement of actual effective *l*. This is a standard problem with vibrating reed measurements. In the present experiments, the relative change in sound velocity and internal friction could not be resolved better than 100 ppm and 5 %, respectively. To highlight the changes that occur in $Q^{-1}(T)$ near the phase transitions and in the mixed-phase regime, the $Q^{-1}$ data taken at different temperatures have been normalized to the room temperature value $Q^{-1}(T = 293.7\,K) = 6.5\,x10^{-3}$.

After compensating for the earth's magnetic field, ac susceptibility, $\chi_{ac}(T)$, was measured at a driving magnetic field of rms amplitude $h_{ac} = 1Oe$ and frequency $\nu = 111Hz$ over the temperature range 1.7 *K* to 300 *K*. The temperature variations of the real ($\chi_0'$) and imaginary ($\chi_0''$) components of ac susceptibility in the cooling (from 300 *K* to 1.7 *K*) and heating (from 1.7 *K* to 300 *K*) cycles are depicted in figure 1. The thermal hysteresis in $\chi_0'(T)$ and $\chi_0''(T)$, extending from $T_{M\,e} \cong 150K$ to $T_{Ae} \cong 280K$, is an experimental signature of the thermoelastic martensitic transformation. The characteristic temperature $T_{M\,b} \cong 260K$ ($T_{Ab} \cong 170K$) for the beginning of the growth of martensite (austenite) phase at the expense of austenite (martensite) phase while cooling (heating) corresponds to the temperature at which the cooling and heating curves bifurcate (the temperature at which $\chi_0'(T)$ and $\chi_0''(T)$ exhibit a pronounced shoulder in the heating curve). These characteristic temperatures are in excellent agreement with those previously[1] determined from $\rho(T)$ and ND data. The temperature range over which the martensite and austenite phases coexist extends from $T_{Mb}$ to $T_{Me}$ on the cooling $\chi_0'(T)$ or $\chi_0''(T)$ curve and from $T_{Ab}$ to $T_{Ae}$ on the heating curve. The Curie temperature, $T_C$, is identified with the temperature corresponding to the dip in the temperature derivative of $\chi_0'(T)$, $d\chi_0'(T)/dT$, versus T curves. This 'inflection-point' (in $\chi_0'(T)$) method yields $T_C = 223K$ ($T_C = 225K$) in the cooling (heating) run.

Figure 2 displays the temperature variations of $\delta V/V$ and $Q^{-1}$ for *q*-$Fe_{20}$, measured at a drive voltage of 7.9 V, when the sample is cooled down to 190 *K*. As the sample temperature decreases, $\delta V/V$ declines while $Q^{-1}$ increases. This decline (rise) in $\delta V/V$ ($Q^{-1}$) signals phonon "softening" in the premartensitic regime. The rate of decline in $\delta V/V$ or equivalently, increase in $Q^{-1}$ *slows down* as the on-set temperature for the growth of the martensitic phase, $T_{Mb} \cong 260K$, is approached while cooling. This feature is followed, at $T < T_{Mb}$, by an upturn (downturn) in $\delta V(T)/V$ ($Q^{-1}$(T)) at $T_{up} \cong 236K$ and subsequently by a sharp drop (rise) at $T_{rap} \cong 232K$. The temperature $T_{up}$ ($T_{rap}$) marks the stage within the mixed-phase regime at which the volume fraction of the tetragonal martensitic phase grows at a slow pace (starts growing at a very rapid rate) at the expense of the parent cubic austenite phase while cooling. These values of $T_{up}$ and $T_{rap}$ tally quite well with those corresponding to the upturn and steep rise observed in $\rho(T)$ previously[1] and in $\chi_0'(T)$ or $\chi_0''(T)$ now (Fig.1). The resonance phase-lock is completely lost when the temperature falls below $218K$ so much so that the reed stops oscillating and the matching electrode picks up only spurious signals. An attempt was made to induce vibrations in the reed by increasing the drive amplitude from the normal value of 5V rms to 15 V rms but without success. Since the drive force varies as the square of the driving amplitude, this increase in the drive amplitude corresponds to about 10 times larger driving force. Consistent with our earlier observations[1] based on the $\rho(T)$ and ND results, this "stiffness" of the reed is indicative of a massive martensitic transformation at such temperatures in the $q$-$Fe_{20}$ sample.

To investigate the martensitic transformation (MT) in greater detail, three different experiments covering (a) the premartensitic regime, (b) part of the mixed-phase regime, and (c) the entire temperature range over which MT occurs, have been carried out. In these experiments, before heating the sample back to room temperature, the sample is cooled down to the temperature $T^*$, which lies (a) above $T_{M\,b} \cong 260K$ ($T^* \cong 265K$), (b) well below $T_{M\,b}$ but not too deep into the austenite-martensite phase coexistence region ($T^* = 238.65K \cong T_{up}$), and (c) well below the temperature $T_{M\,e} \cong 150\,K$, which marks the end of the martensitic transformation, ($T^* \cong 130K$). The results of the experiments (a) – (c) are shown in figures 3 – 5, respectively. These results, when viewed in the light of the relation $V = \sqrt{E/\rho}$, where $E$ is the Young's modulus and $\rho$ is the mass density, imply the following. When $T^* > T_{M\,b}$ (case (a), premartensitic regime), $\delta V/V$ and $Q^{-1}$, measured at a drive voltage of 7.9 V, both exhibit thermal hysteresis (Fig.3). While cooling from 294.5 $K$, the parent austenite phase remains

"elastically stiff " in its $T = 294.5K$ state till the temperature 291.5 K is reached when the phonon "softening" starts and this process continues till $T^* \cong 265K$. On the other hand, in the heating cycle, the austenite phase in the "kinetically arrested" metastable state, corresponding to $T = T^*$, persists till the temperature reaches the value $T \cong 275K$ (which is close to the characteristic temperature $T_{Ae} \cong 280\,K$) beyond which the elastic recovery towards the room temperature parent austenite phase gets started. The elastic recovery process is nearly complete at $T = 294.5K$ when the thermoelastic hysteresis tends to vanish. These results assert that in the premartensitic regime, the parent austenite phase undergoes changes similar to those occurring in the high-temperature high-crystallographic-symmetry phase prior to a first order phase transition. When $T^* \cong T_{up} < T_{M\,b}$ (case (b), Fig.4), the sample has barely entered into the two-phase regime and with temperature increasing from $T^*$, the austenite phase does not recover fully to its room temperature elastic state even when $T > T_{Ae}$, instead the system retains partially the "memory" of the elastic state in the two-phase region. Consequently, unlike the major $\delta V / V$ - T and $Q^{-1}$ - T thermal hysteresis loops (Fig.5), the minor $\delta V(T)/V$ and $Q^{-1}(T)$ thermal hysteresis loops, taken at a drive voltage of 7.9 V, do not close at room temperature. By contrast when the sample is cooled down to $T^* < T_{M\,e}$ (case (c), Fig.5) and the thermal cycling interval embraces the entire temperature range over with the MT takes place, $\delta V(T)/V$ and $Q^{-1}(T)$ data, taken at a drive voltage of 2.13 V, exhibit a rich structure (i.e., peaks, minima, change of slope, etc.) symptomatic of the thermoelastic processes occurring in the two-phase region. Note that, as was the case in the previous experimental run (Fig.2), in this run too, the resonance phase-lock is completely lost when the temperature falls below $218K$ during cooling but the resonance phase-lock is automatically restored when the temperature exceeds 218 K in the heating cycle. From the temperatures corresponding to the peaks, minima and change of slope in the $\delta V(T)/V$ and $Q^{-1}$(T) curves marked in Fig.5, the following important observations can be made. (i) The characteristic temperatures $T_{Mb}$, $T_{Ae}$ and $T_C$ are more sharply defined in the sound velocity and internal friction data than in the $\rho(T)$ (reference 1) and $\chi_0'(T)$ or $\chi_0''(T)$ data (Fig.1). (ii) In the *heating* cycle, the system remembers the initiation (marked by the peak in $\delta V(T)/V$ or dip in $Q^{-1}(T)$ at $T = T_{Mb} = 260\,K$) and subsequent growth (represented by the dip in $\delta V(T)/V$ or peak in $Q^{-1}(T)$ at $T = 236\,K$) of the martensitic phase (at the expense of the parent austenite phase) that had occurred during the cooling cycle in the two-phase regime. (iii) In both cooling and heating cycles, $\delta V(T)/V$ ($Q^{-1}(T)$) exhibits pronounced minima (peaks) corresponding to the Curie temperatures $T_C = 223K$ and $T_C = 225K$, obtained from the $d\chi_0'(T)/dT$ data (Fig.1).

In short, we report the elastic anomalies associated with the austenite – martensite phase transitions as they manifest themselves in the sound velocity and attenuation data and establish correlations with the results obtained by other standard but *indirect* measuring techniques such as electrical resistivity and magnetic susceptibility. Detailed vibrating reed measurements on $Ni_{55}Fe_{20}Al_{25}$ samples with varied degree of site disorder and the analyses of the results, so obtained, are in progress.

**FIGURE CAPTIONS**

**FIG. 1. Temperature variations of the real ( ) and imaginary ( ) components of ac susceptibility in the cooling (open circles) and heating (continuous curves) cycles. The characteristic temperatures for the beginning, ( ), and end, ( ), of the growth of martensite (austenite) phase during cooling (heating) are indicated.**

**FIG. 2. The temperature variations of sound velocity, $\delta V/V$, and internal friction, $Q^{-1}$, measured at a drive voltage of 7.9 V, when the sample is cooled down to 190 K. The values of the temperatures marking the characteristic features in these curves are indicated.**

**FIG. 3. The temperature variations of sound velocity, $\delta V/V$, and internal friction, $Q^{-1}$, measured at a drive voltage of 7.9 V, when the sample undergoes thermal cycling between 265 K and 294.5 K in the premartensitic regime. The values of the temperatures marking the characteristic features in these curves are indicated.**

**FIG. 4. The temperature variations of sound velocity, $\delta V/V$, and internal friction, $Q^{-1}$, measured at a drive voltage of 7.9 V, when the sample undergoes thermal cycling between 238.5 K and 293 K. The values of the temperatures marking the characteristic features in these curves are indicated.**

**FIG. 5. The temperature variations of sound velocity, $\delta V/V$, and internal friction, $Q^{-1}$, measured at a drive voltage of 2.13 V, when the sample undergoes thermal cycling between 130 *K* and 296 *K*. The values of the temperatures marking the characteristic features in these curves are indicated.**

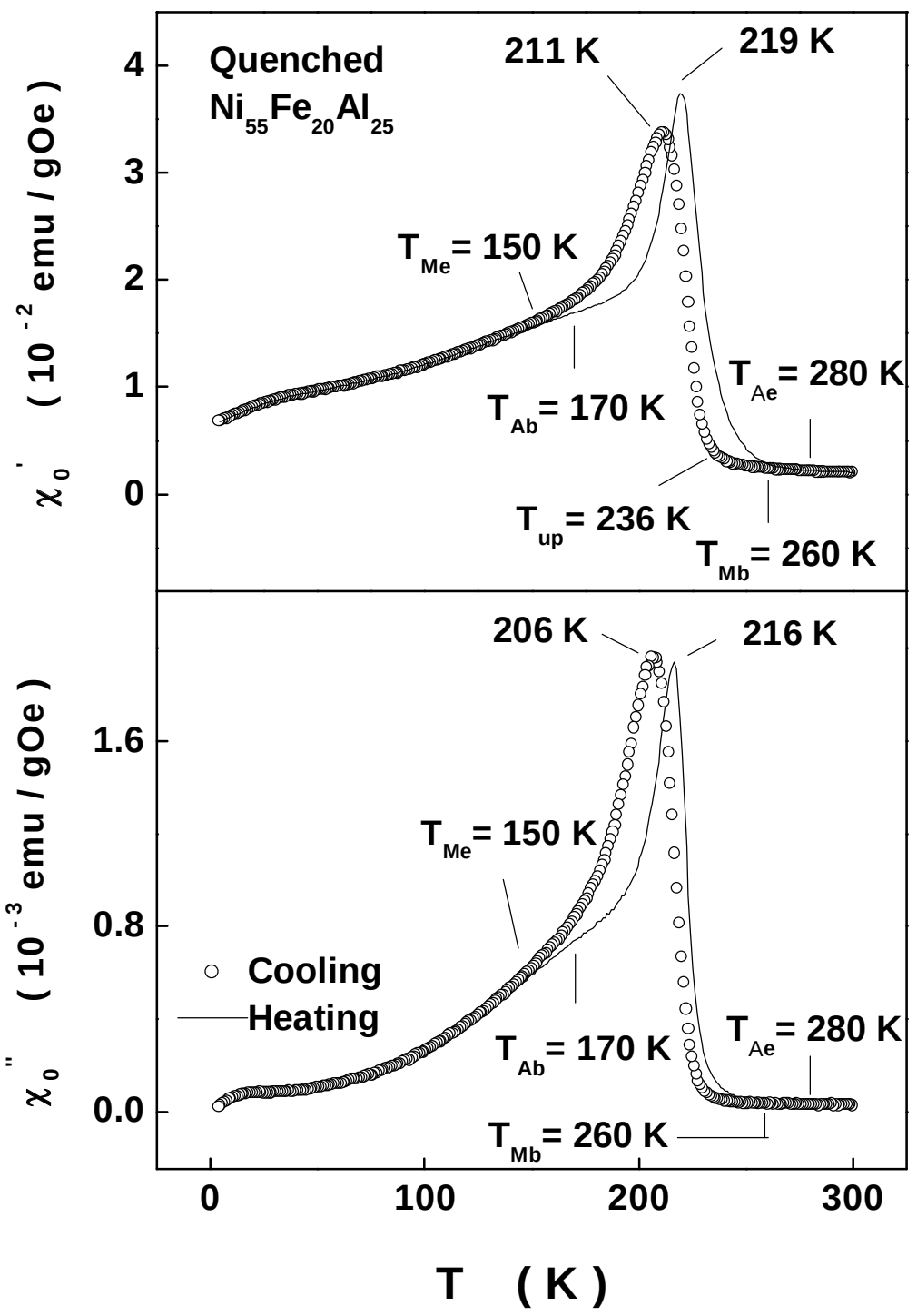

FIG. 1.

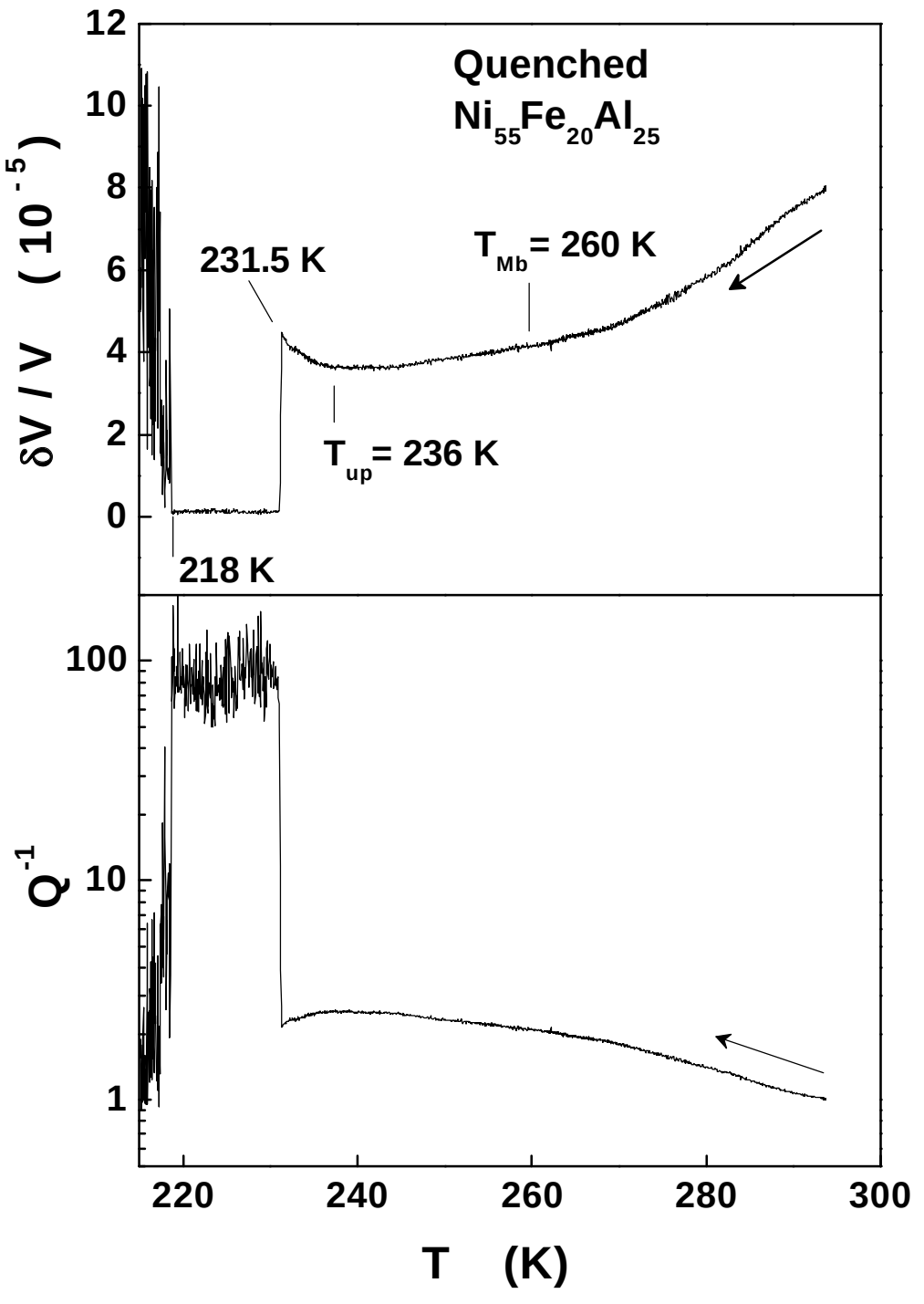

FIG. 2.

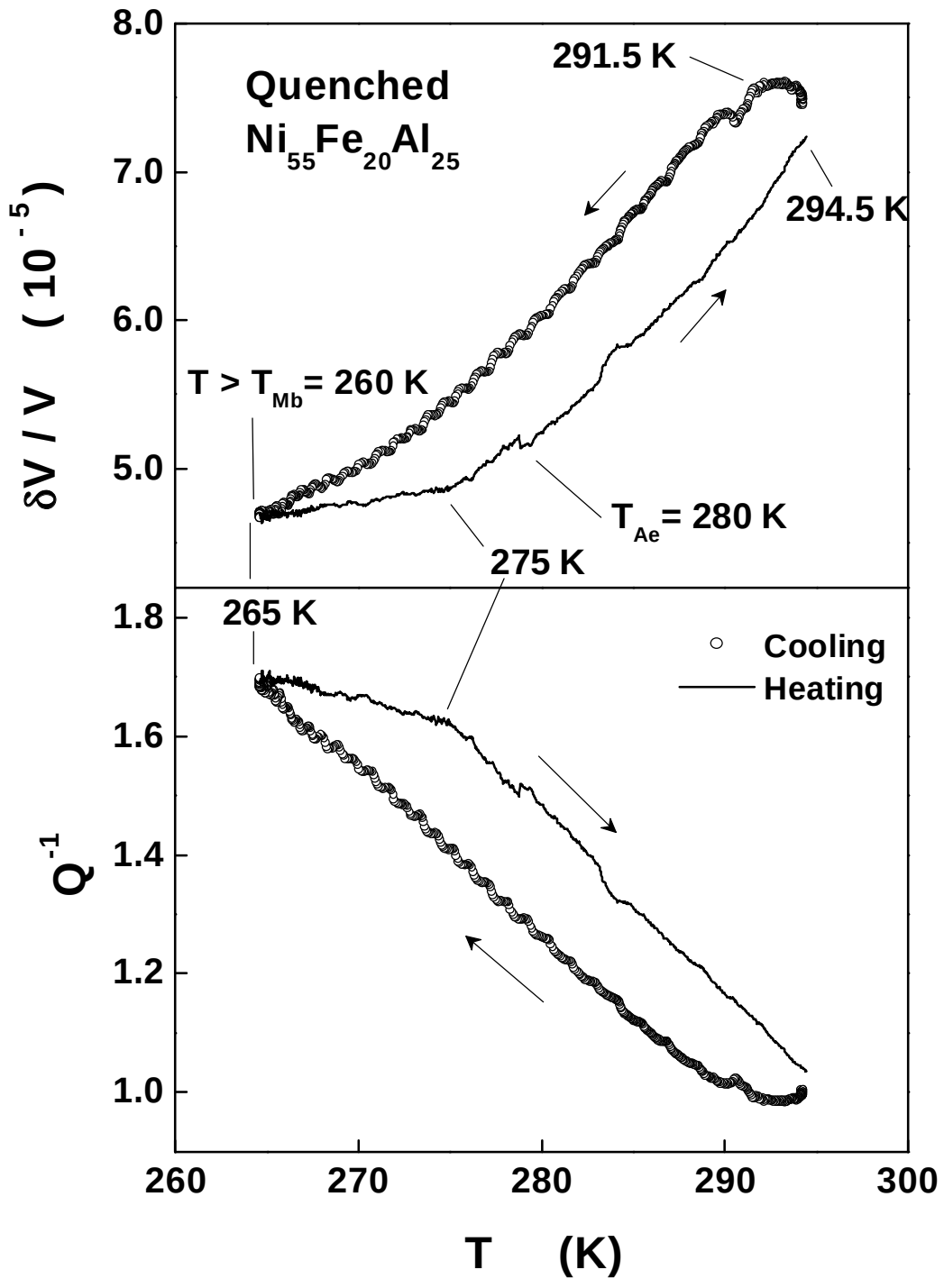

FIG. 3.

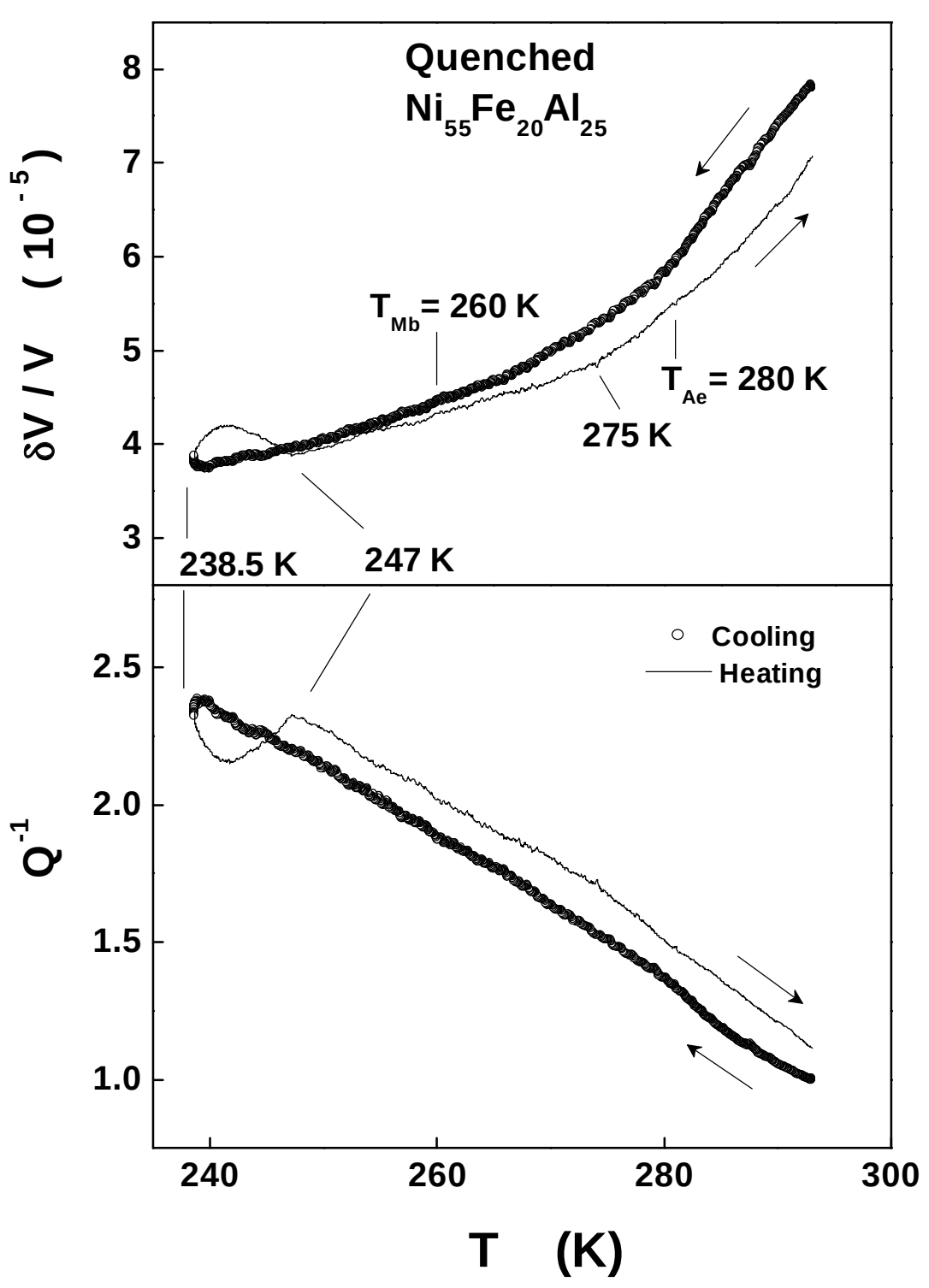

FIG. 4.

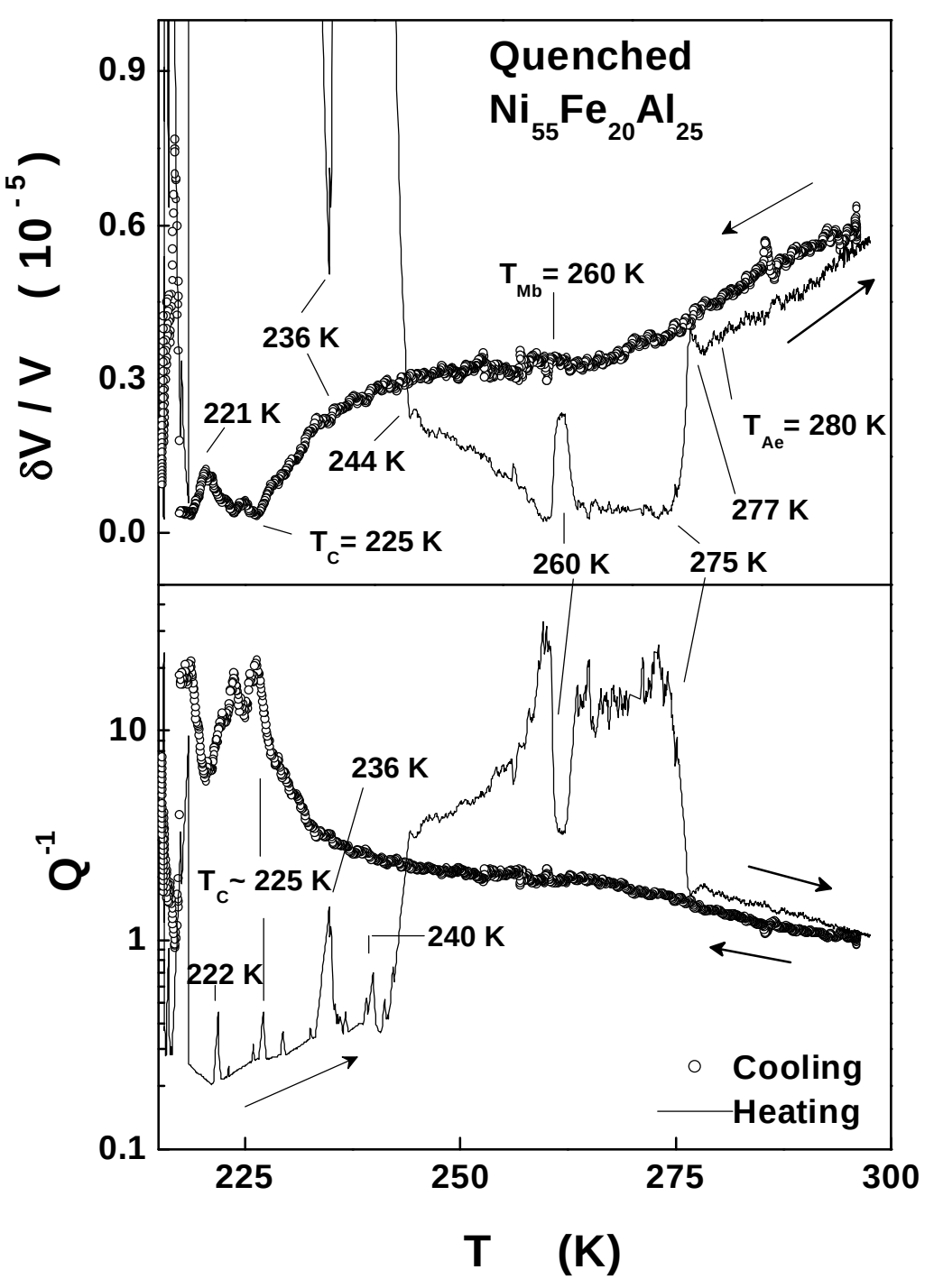

FIG. 5.